\documentclass{ifacconf}
\usepackage{amsmath,amssymb,mathtools}
\usepackage{algorithm}
\usepackage{algpseudocode}

\newtheorem{remark}{Remark}

\usepackage{graphicx}      
\usepackage{natbib}        
\usepackage{braket}
\usepackage{amsmath,amssymb,amsfonts}

\usepackage{graphicx}
\usepackage{textcomp}
\usepackage{xcolor}
\usepackage{subcaption}
\begin{document}
\begin{frontmatter}

\title{Trotterized Variational Quantum Control for Spin-Chain State Transfer\thanksref{footnoteinfo}} 

\thanks[footnoteinfo]{The authors acknowledge the support of the Danish e-Infrastructure Consortium (DeiC) and the National Quantum Algorithm Academy (NQAA) through the Postdoctoral Scholarship under the project ``Quantum-Driven Solutions for Multi-Agent Systems and Advanced Computation''. This work was also partially supported by UID/00147- Research Center for Systems and Technologies (SYSTEC) - and the Associate Laboratory Advanced Production and Intelligent Systems (ARISE, DOI: 10.54499/LA/P/0112/2020) funded by Fundação para a Ciência e a Tecnologia, I.P./ MECI through the national funds.}

\author[First]{Nahid Binandeh Dehaghani}
\author[First]{Rafal Wisniewski}
\author[second]{A. Pedro Aguiar}

\address[First]{Department of Electronic Systems, Aalborg University, Fredrik Bajers vej 7c, DK-9220 Aalborg, Denmark (e-mail: nahidbd@es.aau.dk, raf@es.aau.dk).}
\address[second]{
Research Center for Systems and Technologies (SYSTEC-ARISE), Faculty of Engineering, University of Porto, Rua Dr. Roberto Frias sn, i219, 4200-465 Porto, Portugal, (e-mail: pedro.aguiar@fe.up.pt).}

\begin{abstract}
We present a hybrid variational framework for quantum optimal control aimed at high-fidelity \emph{state transfer} in spin chains. The system dynamics are discretized
and compiled into a parameterized circuit, where deterministic two-qubit blocks implement the drift interactions, while trainable on-site $R_Z$ rotations encode the control inputs.
We study two parameterizations: a compact \emph{global} scheme with a small number of shared parameters per slice, and a \emph{local} scheme with site-wise angles. 
Using a Sequential Least Squares Quadratic Programming (SLSQP) optimization to minimize infidelity, simulations on XXZ spin chains show that both parameterizations can achieve near-unit fidelities in the noiseless regime.
Under depolarizing noise, the global scheme provides improved robustness for comparable circuit depth and iteration budgets. The results make explicit an expressivity--stability trade-off and suggest a scalable route to Noisy Intermediate-Scale Quantum (NISQ) compatible control synthesis.
\end{abstract}


\end{frontmatter}

\section{Introduction}

Reliable manipulation of quantum systems is central to quantum information processing and emerging quantum technologies. Among the available routes to precise state transformations, quantum optimal control \citep{koch2022quantum,dehaghani2023quantumPontryagin,dehaghani2023quantumbalancing, dehaghani2022quantumcdcd} offers a principled framework for synthesizing control fields that steer quantum dynamics toward desired targets while respecting physical and experimental constraints \citep{cerezo2021variational, magann2021pulses}. Recent reviews highlight the growing impact of optimal control across quantum technologies, including computation, simulation, and sensing \citep{yuan2019theory, bharti2022noisy}.

Classical optimal control approaches based on Pontryagin's Maximum Principle (PMP) provide powerful necessary conditions for constrained quantum control problems. However, these methods typically yield coupled two-point boundary value problems that become computationally demanding as system size increases. When control landscapes exhibit stiffness or strong nonlinearities, these formulations require substantial resources and show sensitivity to initialization \citep{dehaghani2023application, chen2025robust, mahesh2023quantum}.

Variational Quantum Algorithms (VQAs) offer a complementary hybrid quantum-classical pathway \citep{li2017hybrid, cerezo2021variational, moll2018quantum}. In this framework, parameterized quantum circuits represent controlled evolution while classical optimizers tune parameters to minimize task-specific objectives. This approach avoids solving computationally expensive adjoint equations required by classical methods, while leveraging the expressivity of parameterized circuits to capture complex quantum dynamics \citep{dehaghani2024hybrid, biamonte2021universal}. Recent advances demonstrate VQAs' feasibility for quantum optimal control, including time-optimal state transfer and robust control under realistic noise models \citep{qi2024variational, xiang2024enhanced, chen2025robust}.

Quantum state transfer in spin chains represents a fundamental protocol for quantum communication and distributed quantum computation. Techniques such as shortcuts to adiabaticity and error mitigation have enabled high-fidelity transfer despite noise and imperfections \citep{xiang2024enhanced, magann2021pulses}, with experimental demonstrations showing scalable protocols in superconducting circuits and other platforms.

Despite these advances, practical quantum control on NISQ devices faces significant challenges from noise, limited coherence times, and hardware constraints. Synthesizing control protocols that achieve optimal state transfer within fixed time horizons while respecting experimental constraints remains a central problem. Developing robust, efficient control strategies for constrained finite-time optimal control therefore represents a key objective for near-term quantum devices.
In this work, we address these challenges through a hybrid variational framework for high-fidelity state transfer in spin chains, systematically comparing global and local control parameterizations and analyzing their robustness under realistic noise models.

\textbf{Contributions}
This paper makes four contributions. (i) We cast quantum state transfer as a variational control problem and give a clean mapping from the bilinear Hamiltonian $H(u(t))$ to a Trotterized circuit with one layer per time slice ($L$), where deterministic two–qubit blocks implement the drift and trainable on-site $R_Z$ rotations encode the controls. (ii) We study two control parameterizations—\emph{global} (shared parameters per slice, $d=2L$) and \emph{local} (site-wise angles, $d=NL$)—and provide depth/parameter counts that make their computational trade-offs explicit. (iii) We specify a practical hybrid loop using SLSQP (finite-difference gradients when no Jacobian is supplied) and a statevector fidelity objective $J(\theta)=1-F(\theta)$, yielding an easily reproducible template for variational control. (iv) Empirically, on XXZ spin chains the \emph{local} scheme attains slightly higher fidelities in the \emph{noiseless} setting, whereas under \emph{depolarizing noise} the \emph{global} scheme is more robust for comparable depth and iteration budgets—revealing a clear expressivity–stability trade-off relevant to near-term implementations.

The remainder of this paper is organized as follows. Section 2 formulates the finite-horizon quantum optimal control problem and establishes its variational representation through Trotterized time discretization. Section 3 details the hybrid variational algorithm, including the control parameterizations, optimization procedure, and implementation considerations. Simulation results on XXZ spin chains are presented and analyzed in Section 3, comparing the performance and robustness of global and local control schemes under noiseless and noisy conditions. Finally, Section 4 concludes the paper and discusses directions for future research.

\section{Optimal Control Formulation}
Consider a finite-dimensional, closed quantum system governed by the bilinear Schrödinger dynamics
\begin{equation}
\begin{aligned}
&i\,\frac{\partial}{\partial t}|\psi(t)\rangle = H(u(t))\,|\psi(t)\rangle,\\
&H(u(t)) = H_d + \sum_{k=1}^{m} u_k(t)\, H_k,
\label{eq:schrodinger}
\end{aligned}
\end{equation}
where \( |\psi(t)\rangle \in \mathbb{C}^n \) denotes the system state, \( H_d \) is the drift Hamiltonian, and \( \{ H_k \}_{k=1}^{m} \) are Hermitian control generators driven by real–valued inputs $u_k(t)$. We set $\hbar=1$ throughout.

\textbf{Admissible controls.} Let the admissible set be
\begin{align*}
 & \mathcal{U} \;=\; \Big\{\, u=(u_1,\dots,u_m): \\ &[0,T]\!\to\!\mathbb{R}^m \;\Big|\;
u_k(\cdot)\ \text{measurable},\ \ u_k^{\min}\le u_k(t)\le u_k^{\max} \Big\}.
\end{align*}
When we later discretize time, we will also consider piecewise–constant controls on a uniform grid.

\textbf{Quantum state transfer.}
Given $|\psi(0)\rangle=|\psi_0\rangle$ and a target $|\psi_f\rangle$, the goal is to steer the system at a fixed horizon $T$ to maximize the terminal fidelity
\[
F(u) \;=\; \big|\langle \psi_f \,|\, \psi_u(T) \rangle\big|^2,
\]
where $|\psi_u(T)\rangle$ denotes the solution of \eqref{eq:schrodinger} under $u(\cdot)$ at time $t=T$.

\textbf{Objective and constraints.}
We pose the control synthesis as a finite-time quantum optimal control problem (OCP) with constraints:
\begin{equation}
    \begin{aligned}
      &\min_{u\in\mathcal{U}} \;\; J(u)\\
\;\;\;&\text{s.t.}\;\;\;
i\,\partial_t|\psi(t)\rangle = H(u(t))|\psi(t)\rangle,\quad
|\psi(0)\rangle=|\psi_0\rangle,
\label{eq:ocp}  
    \end{aligned}
\end{equation}
where the control objective is to maximize the fidelity of state transfer at a fixed terminal time $T$, subject to the system dynamics and admissible control constraints. The terminal cost is defined as
\[
J(u) = 1 - F(u)
\]
where $F(u)$ denotes the fidelity between the final state and the desired target state. Optionally, a regularization term $\lambda\sum_{k=1}^m\int_0^T u_k(t)^2\,dt$ can be added to penalize excessive control effort and enforce physical amplitude bounds.

Problem~\eqref{eq:ocp} thus captures the task of steering the quantum system from an initial state $|\psi_0\rangle$ to a target state within a finite time horizon, while respecting experimental constraints on the control amplitudes. This formulation serves as the foundation for our variational implementation, where in the following, we discretize time 
and map the control inputs 
to circuit parameters, yielding a hybrid quantum--classical solver for~\eqref{eq:ocp}.

\begin{figure}[t]
  \centering
  \includegraphics[width=\linewidth]{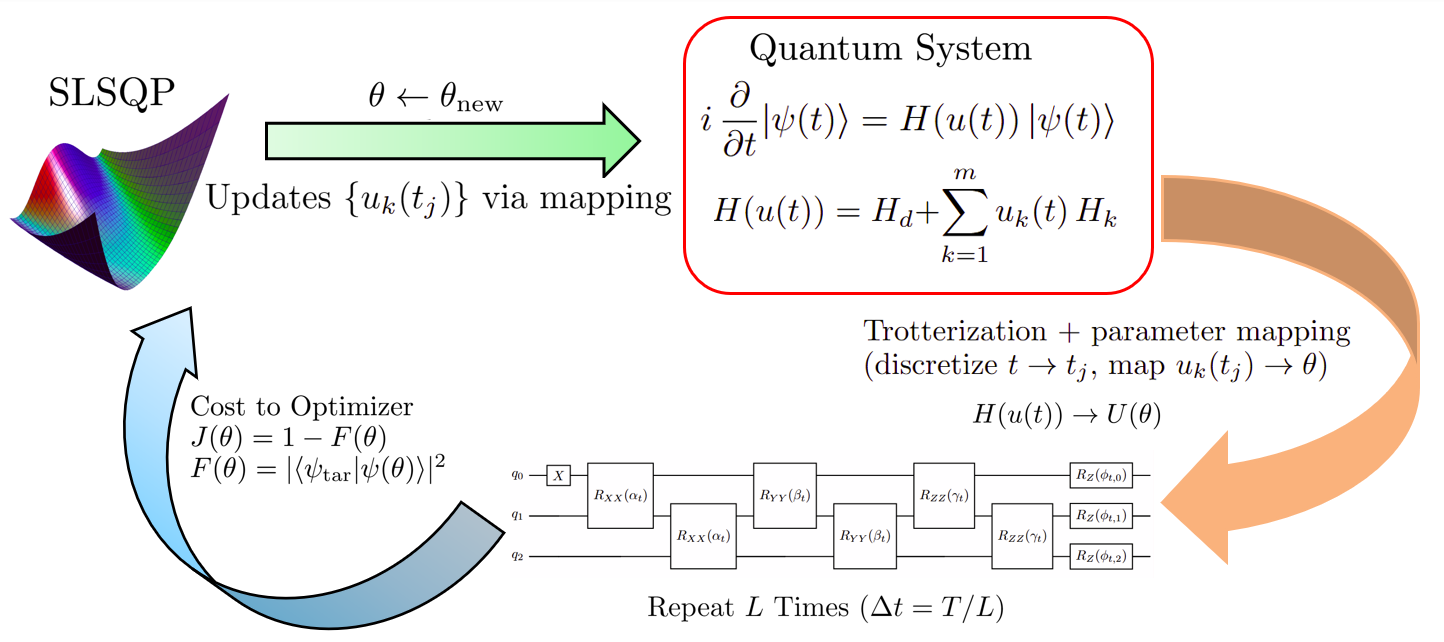}  
  \caption{\small Hybrid variational quantum control loop. The time-discretized Hamiltonian
  is mapped to a parameterized circuit; the simulator returns $F(\theta)$ and
  $J(\theta)=1-F(\theta)$ to SLSQP, which updates $\theta$ and hence $\{u_k(t_j)\}$.
  Repeat $L$ times per pass until convergence.}
  \label{fig:vqa-loop}
\end{figure}

\subsection{Time Discretization and Parametric Representation}
For implementation we discretize $[0,T]$ into $L$ uniform slices with grid points $0=t_0<t_1<\cdots<t_L=T$ and step $\Delta t=T/L$. We adopt piecewise–constant controls on each slice, so the Hamiltonian is held fixed on $[t_\ell,t_{\ell+1})$:
\begin{equation}
H(t)\;\approx\;H\!\left(\boldsymbol{\theta}^{(\ell)}\right),\qquad t\in[t_\ell,t_{\ell+1}),
\label{eq:piecewise}
\end{equation}
where $\boldsymbol{\theta}^{(\ell)}=(\theta^{(\ell)}_1,\ldots,\theta^{(\ell)}_m)$ are tunable parameters obtained by sampling or variationally mapping the controls $\{u_k(t_\ell)\}$.
The slice propagator is
\begin{equation}
U^{(\ell)}\!\left(\boldsymbol{\theta}^{(\ell)}\right)=\exp\!\big(-i\,H(\boldsymbol{\theta}^{(\ell)})\,\Delta t\big),
\label{eq:propagator}
\end{equation}
and the total evolution over $[0,T]$ is the ordered product
\begin{equation}
U(\boldsymbol{\theta}) \;=\; U^{(L)} U^{(L-1)} \cdots U^{(1)},\!\!\!\!
\qquad
\boldsymbol{\theta}=\big\{\boldsymbol{\theta}^{(1)},\ldots,\boldsymbol{\theta}^{(L)}\big\}.
\label{eq:full_evolution}
\end{equation}

When the Hamiltonian admits a decomposition of the form $H(\cdot) = H_d + \sum_k \theta^{(\ell)}_k H_k$, the time-evolution operator for each slice, $U^{(\ell)}$, can be efficiently approximated using a first-order Suzuki–Trotter factorization. This approach separates the contributions from the drift ($H_d$) and control generators ($H_k$), resulting in a quantum circuit layer for each time slice that mirrors the physical interactions and control actions. The full set of parameters $\boldsymbol{\theta}$, encoding the control inputs across all slices, is then optimized within the parameterized quantum circuit to maximize the terminal fidelity (or equivalently, minimize the infidelity). This construction establishes a direct correspondence between the continuous-time optimal control problem and its hybrid quantum–classical solution. The overall workflow is illustrated in Fig.~\ref{fig:vqa-loop}.

\subsection{Quantum Optimal Control Objective}
The goal is to choose parameters $\boldsymbol{\theta}$ that maximize the terminal fidelity with a prescribed target $|\psi_f\rangle$. We minimize the terminal \emph{infidelity}
\begin{equation}
\begin{aligned}
\min_{\boldsymbol{\theta}} \quad & J(\boldsymbol{\theta}) \;=\; 1 - F(\boldsymbol{\theta}), \\
\text{s.t.}\quad & |\psi(T;\boldsymbol{\theta})\rangle \;=\; U(\boldsymbol{\theta})\,|\psi_0\rangle, \\
& F(\boldsymbol{\theta}) \;=\; \big|\langle \psi_f \,|\, \psi(T;\boldsymbol{\theta}) \rangle\big|^2,
\end{aligned}
\label{eq:cost}
\end{equation}
with $F(\boldsymbol{\theta})\in[0,1]$. 

\textbf{Variational realization.}
The parameterized evolution is implemented as a quantum circuit with one layer per time slice: each layer realizes
$U^{(\ell)}(\boldsymbol{\theta}^{(\ell)})=\exp\!\big(-i\,H(\boldsymbol{\theta}^{(\ell)})\,\Delta t\big)$
via a Trotterized sequence of gates reflecting the drift and control generators. The full propagator is
$U(\boldsymbol{\theta})=U^{(L)}\cdots U^{(1)}$.

\textbf{Hybrid optimization loop:}
\begin{enumerate}
\item \textbf{Evaluation.} A quantum processor (or statevector simulator) prepares $|\psi(T;\boldsymbol{\theta})\rangle$ and returns the objective $J(\boldsymbol{\theta})=1-F(\boldsymbol{\theta})$; in the simulator we compute $F$ from an overlap, while hardware would estimate $F$ from measurement statistics (or via a basis-change/overlap circuit).
\item \textbf{Update.} A classical optimizer (SLSQP in our experiments) updates $\boldsymbol{\theta}$ using the returned objective (with finite-difference gradients when no Jacobian is supplied).
\item \textbf{Repeat.} Steps 1–2 iterate until a stopping criterion is met (tolerance on $J$, gradient norm, or iteration budget), yielding $\boldsymbol{\theta}^\star$ and high terminal fidelity.
\end{enumerate}

This closed-loop variational scheme provides a direct bridge from the continuous-time control objective to a trainable circuit, enabling high-fidelity state transfer with a depth that scales linearly in the number of time slices $L$.

\subsection{Variational Quantum Optimization Algorithm}
We obtain the control signals via a hybrid quantum--classical routine. A quantum simulator realizes the dynamics with a parameterized circuit, and a classical optimizer updates the parameters to minimize a terminal infidelity. The fidelity is
$F(\boldsymbol{\theta}) \;=\; \big|\langle \psi_f \,|\, U(\boldsymbol{\theta})\,|\psi_0\rangle\big|^2,$
where $U(\boldsymbol{\theta})=U^{(L)}\!\cdots U^{(1)}$ is the Trotterized evolution, and the objective is $J(\boldsymbol{\theta})=1-F(\boldsymbol{\theta})$.

\textbf{Parameterizations.}
We consider two choices for the control parameter map per slice $\ell=1,\dots,L$:
\begin{itemize}
  \item Global-control: all sites share a small set of parameters (e.g., two per slice), yielding a compact representation with fewer variables and smoother landscapes.
  \item Local-control: each site has its own parameter(s), increasing expressivity and controllability at the cost of higher dimension and potentially more complex training dynamics.
\end{itemize}

\textbf{Optimization loop}
At iteration $r$, the simulator prepares $|\psi(T;\boldsymbol{\theta}^{(r)})\rangle = U(\boldsymbol{\theta}^{(r)})|\psi_0\rangle$ and returns
$F^{(r)}$ (via statevector overlap) and $J^{(r)}=1-F^{(r)}$. The classical optimizer updates
$\boldsymbol{\theta}^{(r+1)}$. We use SLSQP (SciPy); when no Jacobian is provided, SLSQP forms finite-difference gradients internally.

\begin{remark}[Ansatz design and expressivity]
Each circuit layer implements one time slice $U^{(\ell)}(\boldsymbol{\theta}^{(\ell)})=\exp\!\big(-i\,H(\boldsymbol{\theta}^{(\ell)})\,\Delta t\big)$ with $\Delta t=T/L$, , so the total circuit depth scales linearly with \( L \). Global-control shares parameters across sites, yielding a compact representation that tends to be more stable at shallow depth. In contrast, local control assigns site-wise parameters, increasing expressivity and controllability at the cost of sharper optimization landscapes. Selecting \( L \) and the parameterization thus involves a trade-off between depth, controllability, and robustness. Notably, global control may act as a form of implicit regularization, improving resilience under noise—this anticipates the robustness results discussed in Section~3.2.
\end{remark}

\begin{algorithm}[t]
\caption{Hybrid Variational Quantum Control (VQC) Loop}
\label{alg:VQCO}
\begin{algorithmic}[1]
\State \textbf{Input:} $|\psi_0\rangle$, $|\psi_f\rangle$, max iterations $R_{\max}$, tolerance $\varepsilon$
\State Initialize parameters $\boldsymbol{\theta}^{(0)}$ (e.g., random)
\For{$r=0$ to $R_{\max}-1$}
  \State $|\psi(T;\boldsymbol{\theta}^{(r)})\rangle \leftarrow U(\boldsymbol{\theta}^{(r)})|\psi_0\rangle$
  \State $F^{(r)} \leftarrow \big|\langle \psi_f \,|\, \psi(T;\boldsymbol{\theta}^{(r)}) \rangle\big|^2$
  \State $J^{(r)} \leftarrow 1 - F^{(r)}$
  \State $\boldsymbol{\theta}^{(r+1)} \leftarrow \mathrm{SLSQP}\big(\boldsymbol{\theta}^{(r)};\ J(\cdot)\big)$ \Comment{finite-diff $\nabla J$ if no Jacobian}
  \If{$|J^{(r+1)}-J^{(r)}| < \varepsilon$ \textbf{ or } $J^{(r+1)}<\varepsilon$}
    \State \textbf{break}
  \EndIf
\EndFor
\State \textbf{Output:} $\boldsymbol{\theta}^\star=\boldsymbol{\theta}^{(r+1)}$, $F^\star=1-J^{(r+1)}$
\end{algorithmic}
\end{algorithm}

Prior to presenting numerical results, we establish basic well-posedness (boundedness of $J$, existence of minimizers under compactness, and continuity of $U(\boldsymbol{\theta})$ in the parameters), ensuring the formulation is physically and mathematically sound.

\begin{prop}[Well-posedness]
\label{thm:wellposed}
Let the admissible parameter set $\Theta\subset\mathbb{R}^p$ be nonempty and compact, and assume the cost
$J(\boldsymbol{\theta})=1-F(\boldsymbol{\theta})$ is continuous on $\Theta$, where
$F(\boldsymbol{\theta}) = \big|\langle \psi_f \,|\, U(\boldsymbol{\theta})|\psi_0\rangle\big|^2$.
Then:
\begin{enumerate}
\item Boundedness: $0 \le J(\boldsymbol{\theta}) \le 1$ for all $\boldsymbol{\theta}\in\Theta$.
\item Existence of a minimizer: $J$ attains a global minimum on $\Theta$.
\item Physical realizability: $U(\boldsymbol{\theta})$ is unitary for all $\boldsymbol{\theta}$.
\item Convergence to stationarity (algorithmic property): 
Consider any deterministic iterative method that generates a sequence
$\{\boldsymbol{\theta}^{(r)}\}\subset\Theta$ with (i) bounded steps, (ii) sufficient decrease or
descent-type acceptance, and (iii) a standard constraint qualification. Then every accumulation
point of $\{\boldsymbol{\theta}^{(r)}\}$ is a first-order stationary point of $J$ (e.g., satisfies
$\nabla J(\boldsymbol{\theta})=0$ when $J$ is differentiable).
\end{enumerate}
\end{prop}

\begin{pf}
(1) Since $F(\boldsymbol{\theta})$ is the squared modulus of an inner product between normalized
states, $F(\boldsymbol{\theta})\in[0,1]$, hence $J=1-F\in[0,1]$. 
(2) Continuity of $J$ on compact $\Theta$ implies existence of minimizers by the extreme value theorem.
(3) Each slice $U^{(\ell)}(\boldsymbol{\theta}^{(\ell)})=\exp(-i\,H(\boldsymbol{\theta}^{(\ell)})\Delta t)$
is unitary because $H(\boldsymbol{\theta}^{(\ell)})$ is Hermitian; products of unitaries are unitary.
(4) Under the stated assumptions, classical results in nonlinear optimization ensure that any limit
point of a bounded descent sequence is stationary; see~\citep{nocedal2006numerical}. \qed
\end{pf}

\begin{remark}[On assumptions]
Compactness of $\Theta$ is satisfied in practice by imposing box bounds on parameters.
If $J$ is only Lipschitz and piecewise differentiable (e.g., due to finite-difference evaluations),
the statement in item~(4) holds with Clarke stationarity.
\end{remark}

\section{Simulation Results and Analysis}
We consider a one–dimensional spin chain of $N$ qubits with nearest–neighbor interactions and time–dependent local $Z$ fields. The dynamics are
$H(t)=H_d+H_c(t)$ with Heisenberg–type drift
\begin{equation}
H_d \;=\; \sum_{k=1}^{N-1}\!\big(J_x\,X_kX_{k+1} \;+\; J_y\,Y_kY_{k+1} \;+\; J_z\,Z_kZ_{k+1}\big),
\label{eq:drift}
\end{equation}
where $J_x,J_y,J_z$ are coupling strengths and $X_k,Y_k,Z_k$ are Pauli operators on site $k$.
Control acts locally via single–qubit $Z$ terms:
\begin{equation}
H_c(t) \;=\;
\begin{cases}
\displaystyle\sum_{j=1}^{N} u\!\big(C(t), d(t), j\big)\,Z_j, & \text{(global control)},\\[4pt]
\displaystyle\sum_{j=1}^{N} u_j(t)\,Z_j, & \text{(local control),}
\end{cases}
\label{eq:control}
\end{equation}
with the global (harmonic) profile
\begin{equation}
u\!\big(C,d,j\big) \;=\; \tfrac{1}{2}\,C(t)\,\big(j-d(t)\big)^2 .
\label{eq:harmonic}
\end{equation}

\textbf{Task: single–excitation transfer}
We study left–to–right transfer within a fixed horizon $T$:
\[
|\psi_{\mathrm{in}}\rangle = |1\,0\,0\,\cdots\,0\rangle, \qquad
|\psi_{\mathrm{tar}}\rangle = |0\,0\,\cdots\,0\,1\rangle .
\]
We discretize $[0,T]$ into $L$ uniform slices with grid points $t_\ell=\ell\,\Delta t$ for $\ell=0,\dots,L-1$ (and $t_L\equiv T$), where $\Delta t=T/L$. Each slice corresponds to one circuit layer implementing $U^{(\ell)}$. Using first–order Suzuki–Trotter in the same order as our implementation (drift then control),
\begin{equation}
U(T) \;\approx\; \prod_{\ell=1}^{L}
\exp\!\big(-i H_d\,\Delta t\big)\;
\exp\!\big(-i H_c(\boldsymbol{\theta}^{(\ell)})\,\Delta t\big),
\label{eq:first_order}
\end{equation}
where $\boldsymbol{\theta}^{(\ell)}$ encodes the control parameters on slice $\ell$.
The drift factor compiles into sequential two–qubit rotations on each edge:
\begin{equation}
\begin{aligned}
&\exp\!\big(-i H_d\,\Delta t\big) \;\approx\;\\
&\prod_{k=1}^{N-1}
e^{-i J_x X_kX_{k+1}\Delta t}\,
e^{-i J_y Y_kY_{k+1}\Delta t}\,
e^{-i J_z Z_kZ_{k+1}\Delta t},    
\end{aligned}
\label{eq:drift_trotter}
\end{equation}
and the control factor into site–local $R_Z$ gates,
\begin{equation}
\exp\!\big(-i H_c(\boldsymbol{\theta}^{(\ell)})\,\Delta t\big)
\;=\;
\prod_{j=1}^{N} R_Z^{(j)}\!\left(2\,u_j(t_\ell)\,\Delta t\right),
\label{eq:control_trotter}
\end{equation}
with
\[
u_j(t_\ell) \;=\;
\begin{cases}
u\!\big(C(t_\ell), d(t_\ell), j\big), & \text{(global),}\\[2pt]
\theta_{\ell,j}, & \text{(local).}
\end{cases}
\]
Thus each layer comprises $3(N\!-\!1)$ two–qubit rotations (XX, YY, ZZ per edge) followed by $N$ single–qubit $R_Z$ gates.

\textbf{Optimization setup.}
We minimize the infidelity $J(\boldsymbol{\theta})=1-F(\boldsymbol{\theta})$ with SLSQP. When no Jacobian is supplied, SLSQP forms finite–difference gradients internally. Box bounds enforce physical limits:
\[
\begin{aligned}
 &\text{Global:}\quad d(t_\ell)\in[\mathrm{di}{-}1,\ \mathrm{df}{+}1],\ \ C(t_\ell)\in[-3,\ 3] \\
 &\text{Local:}\quad u_j(t_\ell)\in[-2\pi,\ 2\pi],\ \ \forall j.
\end{aligned}
\]
Initial parameters are random; runs stop at a tolerance or at $R_{\max}$ iterations.

\begin{remark}[Parameter counts and anchoring]
Local control assigns an independent parameter to each site on each slice, giving $d=NL$ variables. Global control optimizes the two time series $C(t_\ell)$ and $d(t_\ell)$; in our implementation we anchor $d(t_0)=\mathrm{di}$ and $d(t_{L-1})=\mathrm{df}$, so the free global parameters are $(L-2)$ values for $d$ plus $L$ values for $C$, i.e., $d=2L-2$.
\end{remark}

\begin{prop}[Computational complexity]
\label{thm:complexity}
Let $d$ be the number of real parameters, and let $C_{\mathrm{sim}}$ denote the cost of one fidelity evaluation for fixed \( \boldsymbol{\theta} \). Then, in the finite-difference setting, each iteration of SLSQP incurs a computational cost of \( \mathcal{O}(d C_{\text{sim}}) \)  in the finite–difference setting. In particular, $d=NL$ for local control and $d=2L-2$ for global control with endpoint anchoring.
\end{prop}
\begin{pf}
Each iteration of SLSQP evaluates the objective \( J(\boldsymbol{\theta}) \) and approximates the gradient via finite differences, requiring \( \mathcal{O}(d) \) additional function evaluations in the worst case. Since each evaluation incurs cost \( C_{\text{sim}} \), the total cost per iteration is \( \mathcal{O}(d C_{\text{sim}}) \).
\end{pf}

\subsection{Simulation Setup and Convergence Analysis}
\begin{figure}
    \centering
    \includegraphics[width=1\linewidth]{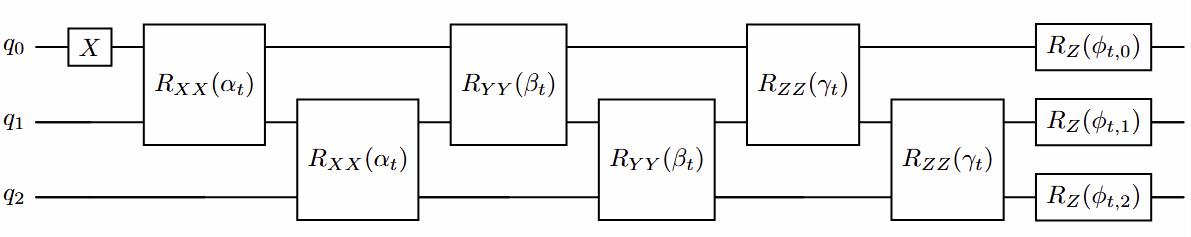}
    \caption{\small One Trotter layer \(t\) for a 3-qubit chain used in the simulations.
Nearest-neighbor drift via \(R_{XX},R_{YY},R_{ZZ}\) followed by on-site \(R_Z\) controls.
Local control: \(\phi_{t,j}=\Delta t\,\theta_{j,t}\).
Global control: \(\phi_{t,j}=\Delta t\,\tfrac12 C_t(j-d_t)^2\).
Stacking \(L\) layers realizes the full evolution.}
\label{fig:three-qubit-layer}
\end{figure}

We simulate a noiseless XXZ spin chain using a statevector backend (\texttt{mqvector}) and evaluate fidelity by overlap (no measurements). Unless stated otherwise, we fix
$N=3,\quad T=2,\quad L=8,\quad  J_x=1,\quad J_y=1,\quad J_z=0.2$
Figure~\ref{fig:three-qubit-layer} shows one Trotter layer used in our simulations (stacked \(L\) times).

The initial and target states are $|100\rangle$ and $|001\rangle$, respectively. For \emph{global} control we optimize two time series $C(t_\ell)$ and $d(t_\ell)$ with anchored endpoints $d(t_0)=\mathrm{di}=0$ and $d(t_{L-1})=\mathrm{df}=N{-}1$; for \emph{local} control we optimize site-wise angles $\theta_{\ell,j}$ so that $u_j(t_\ell)=\theta_{\ell,j}$.

\paragraph*{Parameter initialization and bounds}
Global variables are initialized by a linear ramp for $d(t_\ell)$ from $\mathrm{di}$ to $\mathrm{df}$ and random $C(t_\ell)\!\sim\!\mathcal{U}(-0.5,0.5)$; local variables are initialized i.i.d.\ $\mathcal{U}(-0.5,0.5)$. We impose box constraints matching the implementation:
\[
\text{Global:}\quad d(t_\ell)\in[\mathrm{di}{-}1,\mathrm{df}{+}1],\ \ C(t_\ell)\in[-3,3];
\]
\[
\text{Local:}\quad u_j(t_\ell)\in[-2\pi,2\pi].
\]
We minimize $J(\boldsymbol{\theta})=1-F(\boldsymbol{\theta})$ using SLSQP  with tolerance $10^{-4}$ and a maximum of $R_{\max}$ iterations. Since no Jacobian is supplied, SLSQP forms finite-difference gradients. Let $J^{(q)}$ denote the objective value at the $q$-th \emph{function evaluation}; because finite differences require additional calls, $q$ is larger than the reported SLSQP iteration count. We terminate when either $|J^{(q+1)}-J^{(q)}|<10^{-4}$ or the iteration budget is reached.

\begin{figure}[t]
  \centering
  \includegraphics[width=\linewidth]{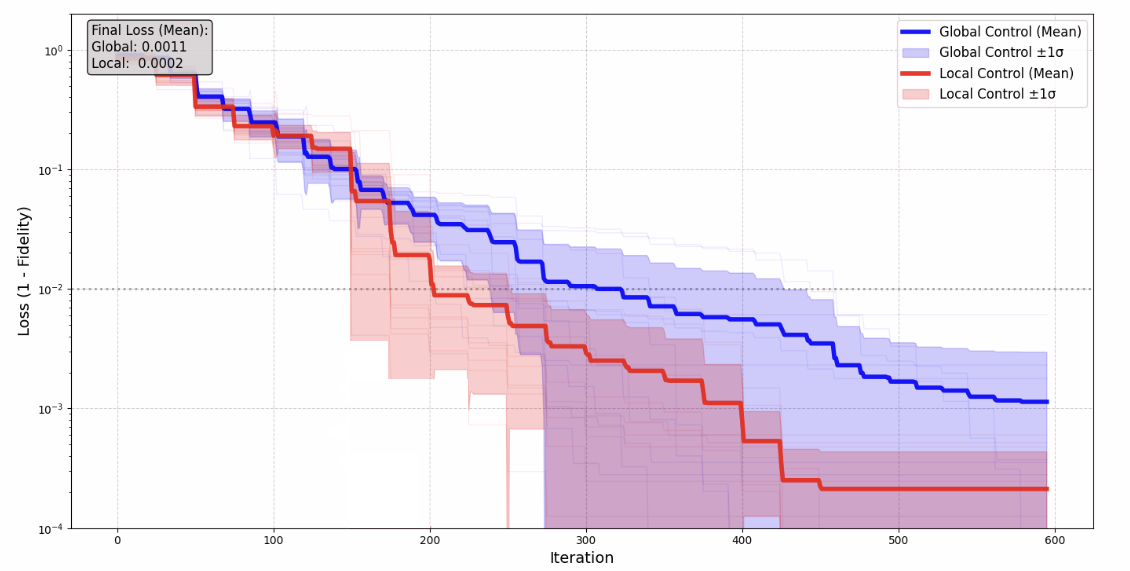}
  \caption{\small Optimization trajectories (semilog scale) for global and local control across 10 realizations ($N=3$, $T=2$, $L=8$). Solid lines: mean loss $J(\theta)$; shaded bands: $\pm1\sigma$. The dashed line marks $J=10^{-2}$. Local control converges faster and attains lower final loss (mean $\sim 2\!\times\!10^{-4}$) than global (mean $\sim 1.1\!\times\!10^{-3}$).}
  \label{fig:loss}
\end{figure}

\textbf{Convergence metrics}
Let $\bar{J}(q)$ be the mean loss across realizations at evaluation $q$, and $\sigma_J(q)$ its standard deviation. We use $\bar{J}_{\text{final}}=\bar{J}(q_{\max})$ for final loss and 
the smallest $q$ such that $\bar{J}(q)<10^{-2}$
for time-to-threshold.
For the configuration above, 
both parameterizations converge monotonically on average. In the noiseless setting, the \emph{local} scheme achieves a lower final loss (mean $\approx 2\times10^{-4}$) than the \emph{global} scheme (mean $\approx 1.1\times10^{-3}$). The local parameterization also reaches the $10^{-2}$ threshold earlier (about $200$ objective evaluations versus $\sim300$ for global) and exhibits tighter confidence bands beyond $\sim250$ evaluations, indicating more stable late-stage refinement. Figure~\ref{fig:loss} summarizes these trajectories (mean $\pm1\sigma$ on a semilog scale) plotted over $50$ realizations.

To assess the time budget required for reliable state transfer, we sweep the total evolution time \(T\) and, for each value, run multiple independent realizations of the variational optimization. Every realization uses a different random initialization of the variational parameters (fixed base seed plus offset), and SLSQP is employed to minimize the infidelity \(J=1-F\). We report the \emph{final} fidelity (after optimization) for each run and aggregate across realizations by plotting the mean \(\pm\) one standard deviation.

\begin{figure}[t]
  \centering
  \includegraphics[width=\linewidth]{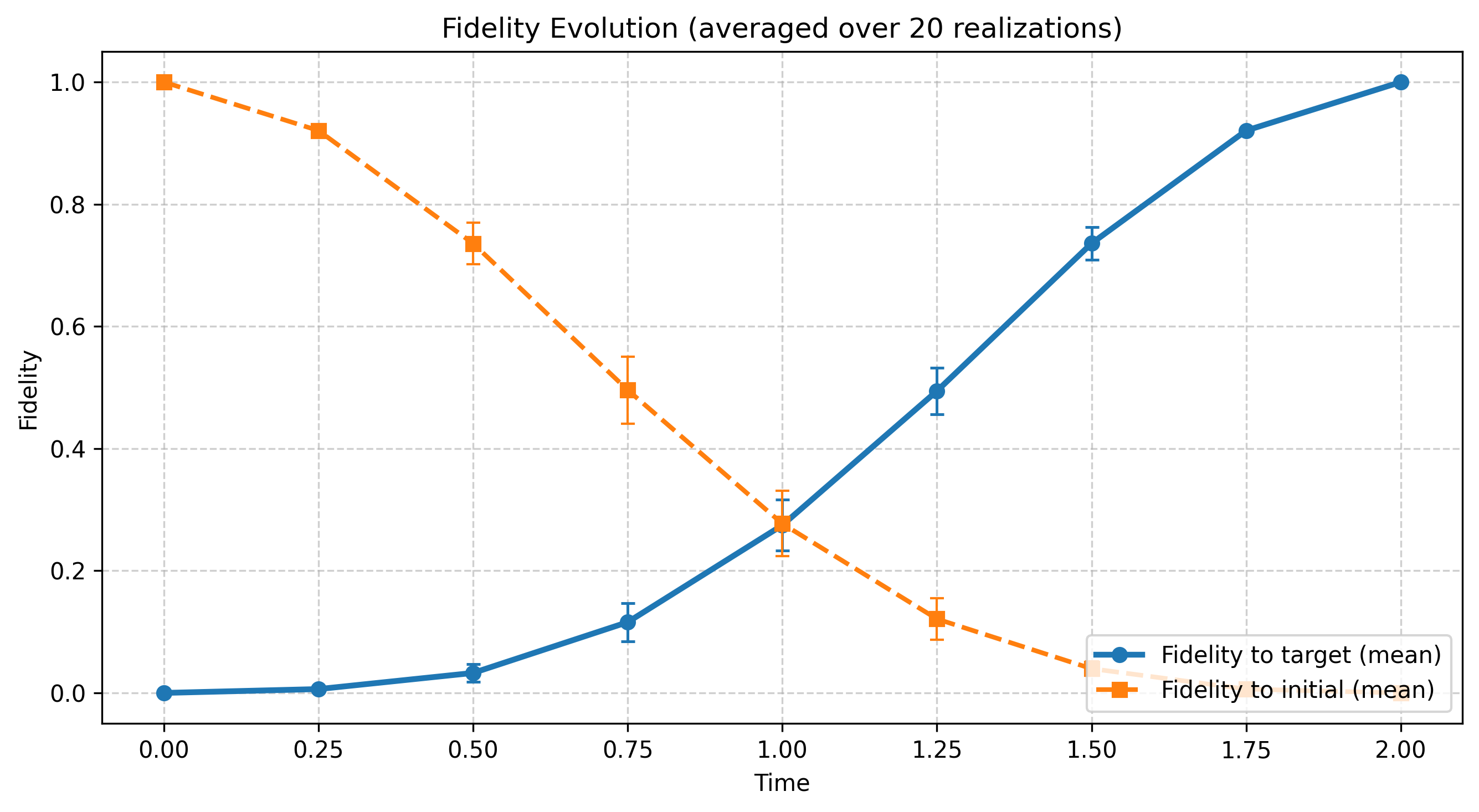}
  \caption{\small Average fidelity evolution for state transfer in a \(N=3\) XXZ chain
  (\(J_x=J_y=1,\,J_z=0.2\)) over a total time \(T=2.0\) discretized into
  \(L=8\) layers (\(\Delta t=0.25\)). Solid blue: mean fidelity to the target
  state; dashed orange: mean fidelity to the initial state. Error bars denote
  \(\pm 1\sigma\) across \(R\) independent realizations (different random
  initializations of the control parameters).}
  \label{fig:fidelity-avg}
\end{figure}

\textbf{Averaged fidelity dynamics (local control).}
We evaluate the state-transfer fidelity along the piecewise-constant evolution by recording
\(F_\mathrm{tar}(t_\ell)=|\langle \psi_{\mathrm{tar}}| \psi(t_\ell)\rangle|^2\) and
\(F_\mathrm{in}(t_\ell)=|\langle \psi_{\mathrm{in}}| \psi(t_\ell)\rangle|^2\) at the layer boundaries
\(t_\ell=\ell\,\Delta t\) for \(\ell=0,\dots,L\).
For the \emph{local} ansatz, each qubit has an independent on-site control
\(u_j(t_\ell)=\theta_{\ell,j}\) with bounds \(\theta_{\ell,j}\in[-4\pi,4\pi]\).
We run \(R=20\) independent realizations, each started from a different random initialization
\(\theta_{\ell,j}\sim\mathcal{U}[-3,3]\) (seeded as \(\texttt{SEED}+r\)), and optimize with SLSQP.
The curves in Fig.~\ref{fig:fidelity-avg} show the mean fidelity across realizations at each
\(t_\ell\); vertical bars depict the corresponding standard deviation (\(\pm 1\sigma\)).
As expected, \(F_\mathrm{tar}(t)\) increases while \(F_\mathrm{in}(t)\) decreases, illustrating a
smooth transfer of the single excitation from \(|100\rangle\) to \(|001\rangle\).
The spread (error bars) is largest at intermediate times, reflecting variability in transient
paths induced by different initial controls, and tightens as the optimizer concentrates mass near
the target at \(t=T\).

\begin{figure}[t]
  \centering
  \includegraphics[width=\linewidth]{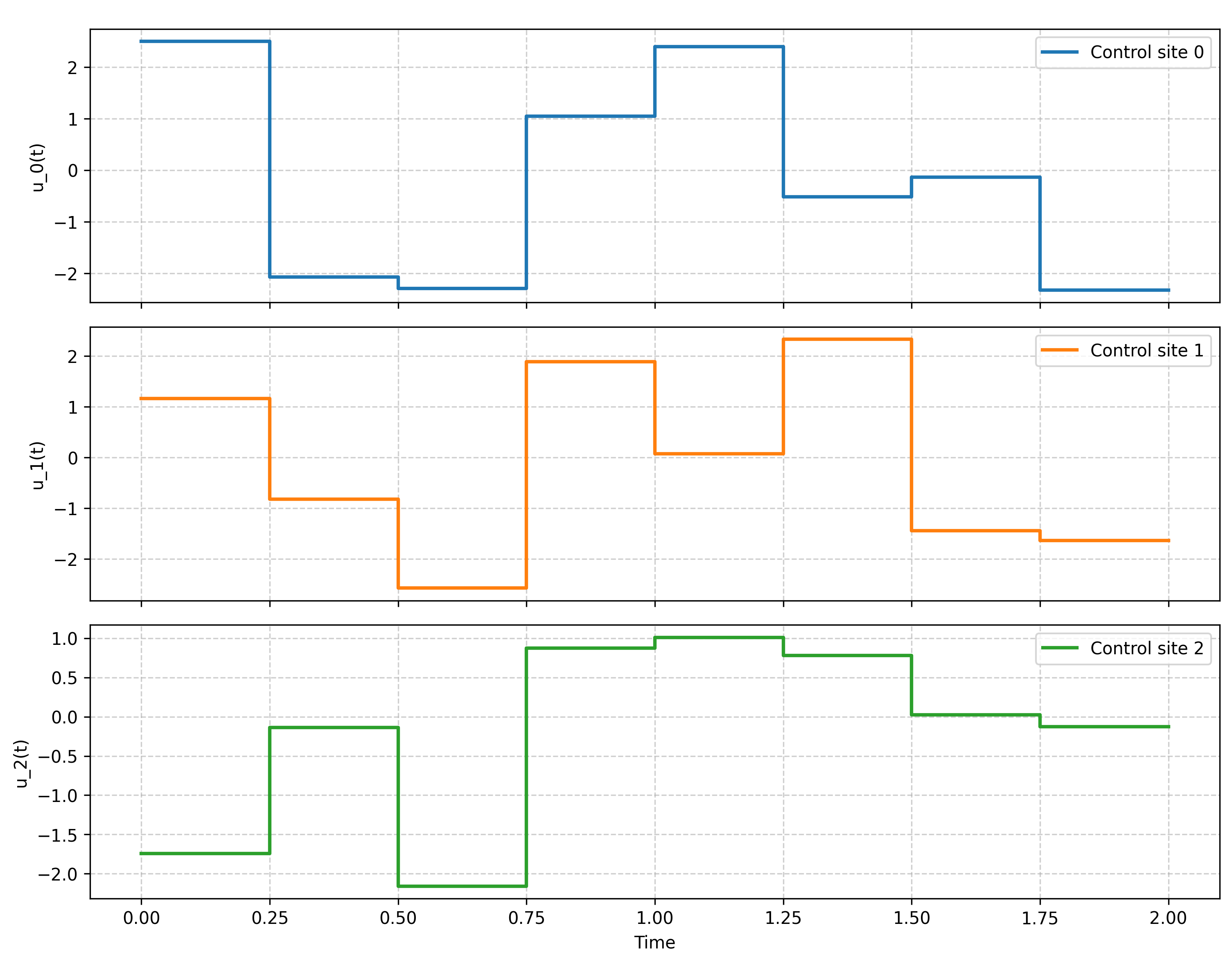}
  \caption{\small Optimized piecewise-constant \emph{local} controls for the best realization
  (highest terminal fidelity) in an \(N=3\) XXZ chain with \(J_x{=}J_y{=}1\), \(J_z{=}0.2\),
  total time \(T{=}2.0\), and \(L{=}8\) layers (\(\Delta t{=}0.25\)).
  Each panel shows the on-site control \(u_j(t)=\theta_{\ell,j}\) held constant on
  \([t_\ell,t_{\ell+1})\) for site \(j\in\{0,1,2\}\).
  Angles are the optimized amplitudes used in the \(R_Z(2\,\theta_{\ell,j}\Delta t)\) gates of our
  Trotterized circuit.}
  \label{fig:controls-local}
\end{figure}

\begin{figure}[t]
  \centering
  \includegraphics[width=\linewidth]{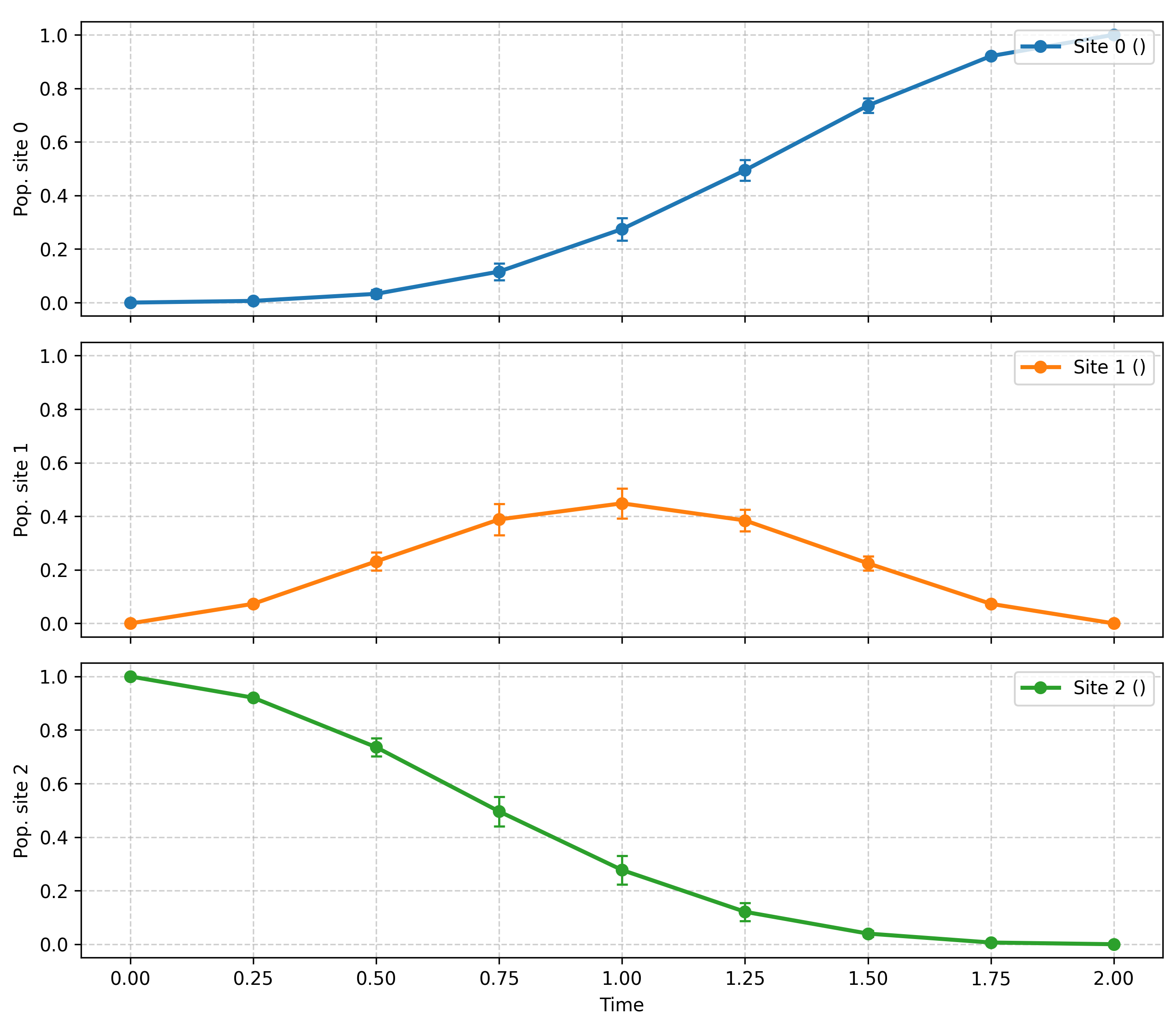}
  \caption{\small Population dynamics at each site of a \(N=3\) XXZ chain
  (\(J_x=J_y=1,\,J_z=0.2\)) during state transfer from \(|100\rangle\) to \(|001\rangle\).
  Simulations use total time \(T=2.0\) discretized into \(L=8\) Trotter layers 
  (\(\Delta t=0.25\)) and \emph{local} piecewise-constant \(R_Z\) controls.
  Curves show the mean populations \(\langle |1\rangle\!\langle 1|_j\rangle\) at sites \(j=0,1,2\)
  versus time, with vertical error bars denoting \(\pm 1\sigma\) across \(R\) independent
  realizations (different random initializations of the control parameters).}
  \label{fig:pop-avg}
\end{figure}

\textbf{Optimized local controls.}
Figure~\ref{fig:controls-local} reports the piecewise-constant on-site fields
for the \emph{best} of \(R\) independent runs (highest terminal fidelity at \(t=T\)).
The pattern is consistent with directional transfer: site \(0\) is driven strongly at early
layers, site \(1\) mediates in the middle, and site \(2\) is activated later to capture the
excitation, in agreement with the averaged fidelity dynamics in Fig.~\ref{fig:fidelity-avg}. Figure~\ref{fig:pop-avg} shows the site populations averaged over \(R\) independent runs of the local-control optimization. The excitation leaves the source site (qubit 2) approximately monotonically, briefly accumulates on the intermediate site (qubit 1), and concentrates on the target site (qubit 0) as \(t \to T\). The vertical error bars (\(\pm 1\sigma\)) quantify the variability across different random initializations. After the transport “turn-on” time, the dispersion shrinks and the final populations approach \((p_0, p_1, p_2) \approx (1, 0, 0)\), consistent with the high terminal fidelity reported in Fig.~\ref{fig:fidelity-avg}.

\textbf{Noise Model and Training Under Noise.}
To evaluate the practical robustness of our variational control framework, we introduce a depolarizing noise model that reflects realistic NISQ device imperfections. The noise is implemented as a depolarizing channel applied after each complete Trotter layer in the quantum circuit.

\textbf{Depolarizing Noise Model and Channel Implementation.}
The depolarizing noise model captures the effect of random Pauli errors that commonly occur in quantum hardware due to environmental interactions and imperfect gate operations. For a single qubit, the depolarizing channel $\mathcal{E}$ acting on a quantum state $\rho$ is defined as
$\mathcal{E}(\rho) = (1-p)\rho + p\frac{I}{2}$,
where $p$ is the depolarizing probability and $I$ is the identity operator \citep{king2003capacity,wilde2013quantum}. This channel represents a quantum state evolving into the maximally mixed state $I/2$ with probability $p$, while remaining unchanged with probability $1-p$.
For multi-qubit systems, we apply independent single-qubit depolarizing channels to each qubit, which for $N$ qubits gives
$\mathcal{E}^{\otimes N}(\rho) = \bigotimes_{i=1}^N \mathcal{E}_i(\rho)$.
In our implementation, we use $p = 10^{-3}$, corresponding to typical error rates in current superconducting quantum processors. The noise channel is applied after each complete Trotter step (each time slice), simulating the cumulative effect of hardware imperfections throughout the quantum state evolution.

\begin{figure}[t]
\centering
\includegraphics[width=0.5\textwidth]{{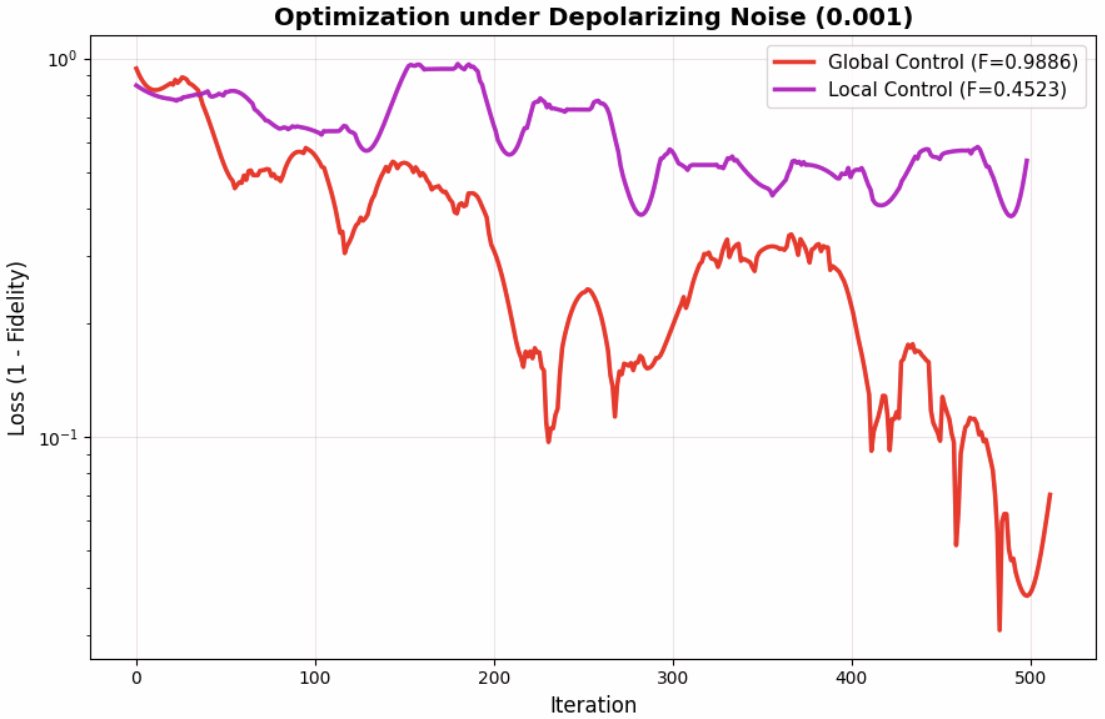}}
\caption{Optimization under depolarizing noise $p = 10^{-3}$. Semi-log loss $\mathcal{L} = 1 - F$ vs. iteration during training: global control reaches $F \approx 0.989$ while local control saturates near $F \approx 0.452$. Shaded region highlights the performance gap.}
\label{fig:noise-optim}
\end{figure}

\textbf{Circuit-Level Noise Implementation}
The depolarizing noise is integrated into our quantum circuit simulation as follows:
\begin{enumerate}
    \item After each complete Trotter layer (comprising drift and control operations)
    \item Applied independently to each qubit in the system
    \item The noise strength $p=10^{-3}$ represents a realistic error rate for current NISQ devices
\end{enumerate}
This implementation models the scenario where each quantum gate operation is followed by a small probability of complete depolarization, making it particularly relevant for near-term quantum hardware where error rates remain significant.
Figure \ref{fig:noise-optim} demonstrates the optimization trajectories under depolarizing noise for both global and local control parameterizations. The results reveal a significant performance gap: the global control strategy consistently achieves high-fidelity state transfer ($F \approx 0.989$), while the local control approach saturates at substantially lower fidelity ($F \approx 0.452$).

This robustness advantage of global control can be attributed to three key factors:
\begin{enumerate}
    \item Parameter Regularization: The global scheme employs only two parameters per time slice ($C_\ell$ and $d_\ell$), compared to $N$ parameters per slice in the local scheme. This reduced parameter space acts as an implicit regularizer, making the optimization less susceptible to noise-induced perturbations.
    \item Spatial Correlation: The harmonic potential profile $u(C,d,j)$ creates spatially correlated control fields that ``sweep'' the excitation through the chain. This correlation helps average out local stochastic errors, providing inherent error suppression.
    \item Smoother Optimization Landscape: The global parameterization yields a more convex and smoother optimization landscape under noise, enabling more reliable convergence compared to the complex, high-dimensional landscape of local control.
\end{enumerate}
The robustness ratio, defined as the fidelity achieved by global control divided by that of local control, demonstrates a $2.19\times$ improvement in noise tolerance for the global parameterization. Notably, both control schemes require similar numbers of optimization iterations, indicating that the performance gap stems from fundamental differences in noise resilience rather than convergence speed.

\section{Conclusion}
We presented a Trotterized variational framework for quantum optimal control of spin-chain state transfer, demonstrating a clear expressivity-stability trade-off between control parameterizations. While local control achieves marginally higher fidelities in noiseless settings, global control exhibits superior robustness under depolarizing noise.
This noise resilience, combined with $\mathcal{O}(L)$ parameter scaling and reduced calibration sensitivity, establishes global control as a practical approach for NISQ applications. The framework provides a scalable pathway from continuous-time control theory to executable quantum circuits, addressing key challenges in near-term quantum device implementation.
Future work will explore larger systems, diverse noise models, and experimental validation, further advancing robust quantum control strategies for emerging quantum technologies.

\bibliography{ref}     

@inproceedings{dehaghani2023quantumbalancing,
  title={Quantum state transfer optimization: Balancing fidelity and energy consumption using pontryagin maximum principle},
  author={Dehaghani, Nahid Binandeh and Aguiar, A Pedro},
  booktitle={2023 IEEE 11th International Conference on Systems and Control (ICSC)},
  pages={94--99},
  year={2023},
  organization={IEEE}
}

@inproceedings{dehaghani2024hybrid,
  title={A hybrid quantum-classical physics-informed neural network architecture for solving quantum optimal control problems},
  author={Dehaghani, Nahid Binandeh and Aguiar, A Pedro and Wisniewski, Rafal},
  booktitle={2024 IEEE International Conference on Quantum Computing and Engineering (QCE)},
  volume={1},
  pages={1378--1386},
  year={2024},
  organization={IEEE}
}

@article{dehaghani2023quantumPontryagin,
  title={Quantum pontryagin neural networks in gamkrelidze form subjected to the purity of quantum channels},
  author={Dehaghani, Nahid Binandeh and Aguiar, A Pedro and Wisniewski, Rafal},
  journal={IEEE Control Systems Letters},
  volume={7},
  pages={2227--2232},
  year={2023},
  publisher={IEEE}
}

@inproceedings{dehaghani2023application,
  title={An application of pontryagin neural networks to solve optimal quantum control problems},
  author={Dehaghani, Nahid Binandeh and Aguiar, A Pedro},
  booktitle={2023 9th International Conference on Control, Decision and Information Technologies (CoDIT)},
  pages={202--207},
  year={2023},
  organization={IEEE}
}

@inproceedings{dehaghani2022quantumcdcd,
  title={A quantum optimal control problem with state constrained preserving coherence},
  author={Dehaghani, Nahid Binandeh and Pereira, Fernando Lobo and Aguiar, Ant{\'o}nio Pedro},
  booktitle={2022 IEEE 61st Conference on Decision and Control (CDC)},
  pages={5831--5836},
  year={2022},
  organization={IEEE}
}

@book{nocedal2006numerical,
  title={Numerical optimization},
  author={Nocedal, Jorge and Wright, Stephen J},
  year={2006},
  publisher={Springer}
}

@article{koch2022quantum,
  title={Quantum optimal control in quantum technologies. Strategic report on current status, visions and goals for research in Europe},
  author={Koch, Christiane P and Boscain, Ugo and Calarco, Tommaso and Dirr, Gunther and Filipp, Stefan and Glaser, Steffen J and Kosloff, Ronnie and Montangero, Simone and Schulte-Herbr{\"u}ggen, Thomas and Sugny, Dominique and others},
  journal={EPJ Quantum Technology},
  volume={9},
  number={1},
  pages={19},
  year={2022},
  publisher={Springer Berlin Heidelberg}
}

@article{li2017hybrid,
  title={Hybrid quantum-classical approach to quantum optimal control},
  author={Li, Jun and Yang, Xiaodong and Peng, Xinhua and Sun, Chang-Pu},
  journal={Physical review letters},
  volume={118},
  number={15},
  pages={150503},
  year={2017},
  publisher={APS}
}

@article{cerezo2021variational,
  title={Variational quantum algorithms},
  author={Cerezo, M. and others},
  journal={Nature Reviews Physics},
  volume={3},
  number={9},
  pages={625--644},
  year={2021}
}

@article{bharti2022noisy,
  title={Noisy intermediate-scale quantum algorithms},
  author={Bharti, K. and others},
  journal={Reviews of Modern Physics},
  volume={94},
  number={1},
  pages={015004},
  year={2022}
}

@article{magann2021pulses,
  title={Pulses and circuits: Two routes to quantum optimal control},
  author={Magann, A. B. and others},
  journal={PRX Quantum},
  volume={2},
  number={1},
  pages={010101},
  year={2021}
}

@article{yuan2019theory,
  title={Theory of variational quantum simulation},
  author={Yuan, X. and others},
  journal={Quantum},
  volume={3},
  pages={191},
  year={2019}
}

@article{chen2025robust,
  title={Robust and optimal control of open quantum systems},
  author={Chen, Z.-J. and others},
  journal={Science Advances},
  volume={11},
  number={43},
  pages={adr0875},
  year={2025}
}

@article{mahesh2023quantum,
  title={Quantum Optimal Control: Practical Aspects and Diverse Methods},
  author={Mahesh, T. S. and others},
  journal={Journal of the Indian Institute of Science},
  volume={103},
  pages={591--607},
  year={2023}
}

@article{moll2018quantum,
  title={Quantum optimization using variational algorithms on near-term quantum devices},
  author={Moll, N. and others},
  journal={Quantum Science and Technology},
  volume={3},
  number={3},
  pages={030503},
  year={2018}
}

@article{biamonte2021universal,
  title={Universal variational quantum computation},
  author={Biamonte, J.},
  journal={Physical Review A},
  volume={103},
  number={L030401},
  year={2021}
}

@article{qi2024variational,
  title={Variational quantum algorithms: fundamental concepts, applications and challenges},
  author={Qi, H. and others},
  journal={Quantum Information Processing},
  volume={23},
  pages={224},
  year={2024}
}

@article{xiang2024enhanced,
  title={Enhanced quantum state transfer by circumventing quantum chaotic behavior},
  author={Xiang, L. and others},
  journal={Nature Communications},
  volume={15},
  pages={48791},
  year={2024}
}

@article{king2003capacity,
  title={The capacity of the quantum depolarizing channel},
  author={King, Christopher},
  journal={IEEE Transactions on Information Theory},
  volume={49},
  number={1},
  pages={221--229},
  year={2003},
  publisher={IEEE}
}

@book{wilde2013quantum,
  title={Quantum information theory},
  author={Wilde, Mark},
  year={2013},
  publisher={Cambridge university press}
}
\end{document}